# Magnetic field induced enlargement of the regime of critical fluctuations in the classical superconductor $V_3Si$ from high-resolution specific heat experiments

Y. Zheng[1], Y. Liu[1], N. Toyota[2] and R. Lortz[1]♣

[1]*Department of Physics, The Hong Kong University of Science and Technology, Clear Water Bay, Kowloon, Hong Kong, China*
[2]*Physics Department, Graduate School of Science, Tohoku University, 980-8571 Sendai, Japan*

**Abstract**
We present high-resolution specific heat data from a high-purity single crystal of the classical superconductor $V_3Si$, which reveal tiny lambda-shape anomalies at the superconducting transition superimposed onto the BCS specific heat jump in magnetic fields of 2 T and higher. The appearance of these anomalies is accompanied by a magnetic-field-induced broadening of the superconducting transition. We demonstrate, using scaling relations predicted by the fluctuation models of the 3d-XY and the 3d-Lowest-Landau-Level (3d-LLL) universality class that the effect of critical fluctuations becomes experimentally observable due to of a magnetic field-induced enlargement of the regime of critical fluctuations. The scaling indicates that a reduction of the effective dimensionality due to the confinement of quasiparticles into low Landau levels is responsible for this effect.

## 1. Introduction

The superconducting transition of classical superconductors in absence of magnetic fields is of second-order nature. Such a transition separates a low-temperature phase, characterized by some form of long-range order, from a phase at higher temperatures in which this order is absent. In superconductors, the long-range order is represented by the macroscopic phase-coherent BCS many-particle wave function of condensed electrons which are bound into Cooper pairs. Second-order phase transitions are usually strongly influenced by fluctuations in the degrees of freedom of the order parameter, represented here by the phase and amplitude of the superconducting wave-function. This is indeed the case for the superconducting transition of high-temperature superconductors (HTSCs) with their high critical transition temperatures and their small coherence volumes [1-8]. As a result, critical fluctuations have a strong impact on most physical quantities. The superconducting transition in zero field is expected to belong to the 3d-XY universality class with a two-component order parameter, in which the phase and amplitude of the Cooper pair wave function represent the two degrees of freedom. All thermodynamic quantities follow therefore characteristic power law dependence with 3d-XY critical exponents. For example, the specific heat will show the typical lambda shape transition anomaly [2,3], which falls in the same universality class as the superfluid transition in liquid helium [9]. The critical temperature is then reduced from the mean-field value by fluctuations in the phase of the order parameter, while Cooper pairs exist in a wide range of temperatures above $T_c$ [4]. To investigate the presence of critical fluctuations, one can try to determine the critical exponents

---
♣ Corresponding author: Lortz@ust.hk

for the thermodynamic data directly from of power law fits. This requires a high sample quality with very sharp phase transition anomalies and a sufficiently large critical temperature regime around $T_c$. Another method is to apply scaling, which means that thermodynamic data is compared to experimental or simulated data from a system that falls into the same universality class [5]. This may require plotting the data as a function of the scaling variables predicted by the corresponding fluctuation model. Due to the large coherence volumes of conventional superconductors and their comparatively low transition temperatures, the range of critical fluctuations in absence of a magnetic field is limited to an extremely small temperature range around the critical temperature $T_c$. Although, it is widely believed that such fluctuations are not observable in bulk classical superconductors, it has been shown in extremely clean superconductors with transition temperatures higher than 10 K that the critical range can be increased in applied magnetic fields and may become experimentally observable in high resolution experiments [10,11,12]. In $Nb_3Sn$ with its high $T_c$ of 18 K, we have previously shown that in a magnetic field of ~0.1 T a small lambda shape anomaly is formed and superimposed on the mean-field like specific heat jump [12]. In parallel, the transition was significantly broadened. Scaling of the specific heat demonstrated that both the lambda shape and the broadening are effects of the field-induced increase in strength of critical fluctuations. The specific heat followed the scaling behaviour of the 3d-XY universality class in small magnetic fields up to ~1.5 T, while in higher fields the expected crossover into the range of 3d 'Lowest Landau Level' fluctuations was observed. To test whether this behaviour is unique for $Nb_3Sn$, we report high-resolution specific heat experiments on $V_3Si$, another member of the superconducting A15 compounds.

## 2. Experiment

The $V_3Si$ sample used in this research is a single crystal of high-quality with $T_c$ = 16.6 K. Its dimensions are 1.0 x 0.3 x 10.0 $mm^3$. The specific heat was measured with an AC heat-flow calorimeter made by a sapphire platform hold by a thermopile of 24 thermocouples. The thermopile is sensitive for the heat flow between the sample and a thermal bath [12]. A resistive Joule heater on the back of the sapphire disc allows us to modulate the sample temperature periodically a frequency of ~1 Hz with amplitude which is kept well below 1 mK to avoid broadening of the phase transition anomalies. A low-noise DC preamplifier is used to pre-amplify the thermocouple signal, which is supplied to a digital lock-in amplifier. The high thermal conductance of the thermopile ensures that the typical DC-offset temperature of the sample platform with respect to the thermal bath is always less than 1 mK and the calorimeter operates basically in an isothermal heat-flux mode. The AC technique offers a very high sensitivity and density of data points (~one data point each 2 mK). For minimizing the broadening effect of the geometric demagnetization factor of the sample, the needle-shaped single crystal was mounted with its long direction along the direction of the applied magnetic field.

## 3. Results

In Figure 1 we show the total specific heat of the sample in various magnetic fields up to 8 T. A small step at $T_M$ = 21.5 K represents the structural martensitic transition [13]. In zero magnetic field, the superconducting transition occurs at $T_c$ = 16.6 K, as indicated by a characteristic BCS mean-field specific heat jump without any fluctuations. In a field of 0.5 T, the magnitude of the jump is reduced by a factor of ~1.2. This is expected for strong type II superconductors with a high-$\kappa$ value ($\kappa = \lambda/\xi$ with the penetration depth $\lambda$ and the coherence length $\xi$) when the transition occurs from the normal state to the Abrikosov state instead of the Meissner state [14]. Finite magnetic fields decrease $T_c$ and broaden the transition significantly in fields of 2 T and above. As shown in the inset, in a field of 1 T, the mean field jump component remains unchanged, while starting from 2 T a tiny peak appears on top of the mean field jump. This small 'lambda' anomaly is present in fields up to 8 T, but gets significantly broadened in higher fields. In this data, the normal state lattice and Sommerfeld contributions are removed for further data analysis. The 8 T data (which was slightly extrapolated below $T_c(8T)$ ) was used as a background for this purpose. This anomaly is similar to what we have observed previously in $Nb_3Sn$, where it could be observed already at much lower fields [12]. Using scaling relations for the specific heat predicted by theory, we have shown that the lambda anomaly together with the broadening of the main step-like transition anomaly at $T_c(H)$ is a consequence of critical fluctuations, which are strongly enhanced by the applied magnetic field. The fact that ~10 times higher magnetic fields are required to achieve the same effect in $V_3Si$ demonstrates that the critical temperature regime of $V_3Si$ is smaller than in $Nb_3Sn$. These scaling relations require the separation of the normal state contribution, as described above, and we will test them in the following to examine the nature of the lambda anomaly further.

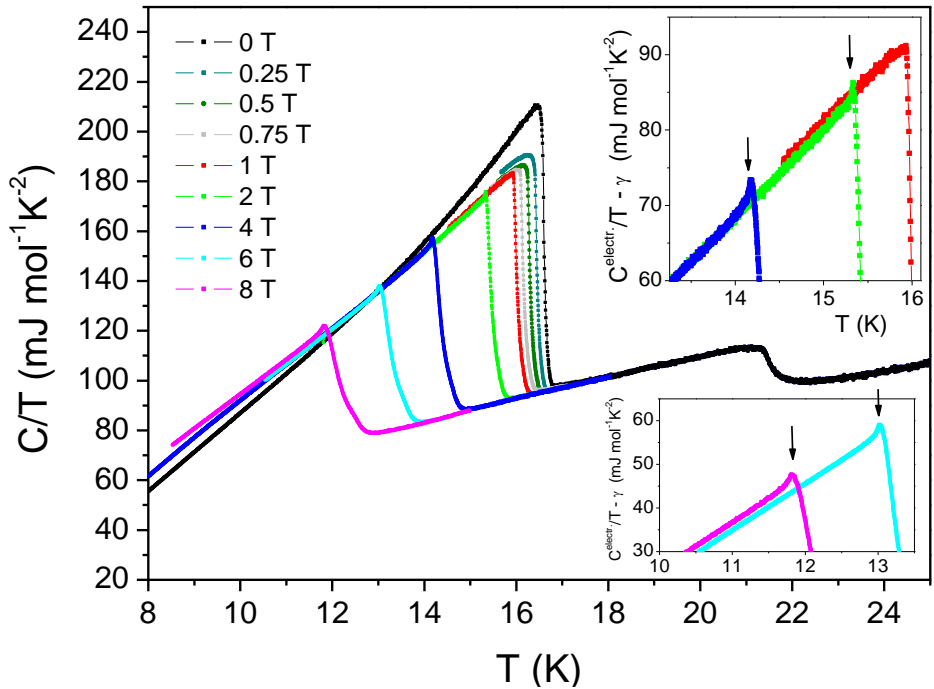

**Figure 1.** Total specific heat of $V_3Si$ in various magnetic fields up to 8 T. The jump-like anomaly at 21 K is the structural martensitic transition, while the superconducting transition occurs at 16.6 K in zero field and is lowered and broadened by the applied magnetic fields. Insets: enlarged view of the tiny 'lambda'-shaped fluctuation anomalies, which appear in fields of 2 T and on top of the broadened mean-field like jump at $T_c(H)$.

The Ginzburg temperature $\tau_G = Gi \cdot T_c$ defines a temperature in the vicinity of $T_c$ at which fluctuation contributions in the specific heat should reach equal magnitude to the mean-field jump. The parameter $Gi$ is known as Ginzburg number and is defined as $Gi = 0.5(k_B T_c)^2/(H_c^2(0)\xi_0^3)^2$ ($H_c(0)$: zero temperature thermodynamic critical field and $\xi_0$: isotropic Ginzburg-Landau coherence length). With $T_c = 16.6$ K, $H_c(0) = 6000$ Oe and $\xi_0 = 40$ Å [15], we obtain $\tau_G = 10^{-6}$ K for V$_3$Si. With our high-resolution experiment, we should be able to resolve much smaller fluctuation contributions down to a few percent of the mean field jump magnitude, which may occur at most in a temperature regime of $10^{-4}$ - $10^{-3}$ K around $T_c$. Our temperature resolution is ~ 1 mK, which is further limited by the finite width of the superconducting transition, which certainly explains the absence of any measureable fluctuation signal in our low field data. However, in a magnetic field $\tau_G$ typically increases dramatically, since the magnetic field confines the quasiparticles in low Landau levels and thus effectively reduces the dimensionality [10,11,16]. To identify the nature of fluctuations causing the small lambda anomaly, a direct way would be fitting the anomaly with power laws of the different fluctuations models to extract the critical exponents. However, the temperature range where this anomaly is visible is too small to obtain reliable information from this. The nature of the fluctuations can be investigated through the increase of the fluctuation regime width in finite fields, which is expected to obey scaling laws with critical exponents. Such scaling of specific heat data in various applied fields successfully described the fluctuation contributions in some HTSCs [2,3,5-7] and in Nb$_3$Sn [12]. The magnetic field introduces a magnetic length $l = (\Phi_0/B)^{1/2}$ that effectively reduces the dimensionality and therefore the coherence volume [5,17]. Scaling means that the data measured in different magnetic fields is normalized by the ratio $\xi/l$ of the coherence length and the magnetic length. If the correct fluctuation model is chosen, scaled data should all be mapped on the same scaling function. 3d-XY scaling was applied successfully to specific heat data of Nb$_3$Sn in fields up to 1 T and of the HTSC YBCO in fields up to 10 T [5]. In the case of Nb$_3$Sn, the confinement of the quasiparticles to low Landau levels caused a failing of 3d-XY scaling in higher fields [12] and a crossover to the 3d-LLL scaling model was observed [6,8,11].

We tested both models on V$_3$Si. In Figure 2 (a) we attempted scaling the low field data according to the 3d-XY model. Since the fluctuation contribution is tiny, we modified the model by normalizing the specific-heat jump $\Delta C$ and used a field dependent $T_c(H)$ instead of a fixed zero field $T_c$ value [12]. If 3d-XY fluctuations were present, the data plotted in the form of $C/\Delta C\ H^{\alpha/2\nu}$ vs. $[T/T_c(H)-1]H^{-1/2\nu}$ should be all mapped on one universal scaling function and thus merge ($\nu \cong 0.669$, $\alpha \cong -0.007$) [1]. The scaling fails for fields up to 1 T. This is not surprising, since the lambda anomaly is not visible in this low field region, indicating the absence of an observable fluctuation contribution. Figure 2 (b) shows the field range between 2 and 8 T, in which the lambda anomalies are appear. Here the scaling is working well, except below the value $[T/T_c(H)-1]H^{-1/2\nu} = -0.005$, which represents the crossover to a mean field behavior in the low temperature regime.

In Figure 2 (c) and (d) we plotted the same data for the two different field regions in the form of $C/\Delta C$ vs. $[T-T_c(H)](HT)^{-2/3}$. This is the scaling normalization predicted by the 3d-LLL fluctuation model. Similar to the 3d-XY model, the 3d-LLL scaling model is applicable in magnetic fields exceeding 1 T, while scaled data in smaller applied fields diverge. The similarity of the scaling diagrams of the two models is not surprising, because the critical exponents that appear in the scaling variables of both models are very similar. This makes it difficult to distinguish the type of fluctuations in higher applied fields of a several Tesla, while differences would appear especially in the lower field range. For $Nb_3Sn$, 3d-XY scaling was limited to fields below 1 T [12]. However, in $V_3Si$ the fluctuations disappear in such small fields. Since these superconductors are quite similar and the 3d-LLL model was applicable in $Nb_3Sn$ in fields above ~2 T, it seems obvious that the magnetic-field-induced dimensionality reduction from the quasiparticle confinement in rather low Landau levels is the main reason for the strong field-induced enlargement of the transition width and the appearance of the small lambda anomalies associated with the fluctuations. The fluctuations in the range of 2 - 8 T are thus associated with fluctuations of the 3d-LLL universality class.

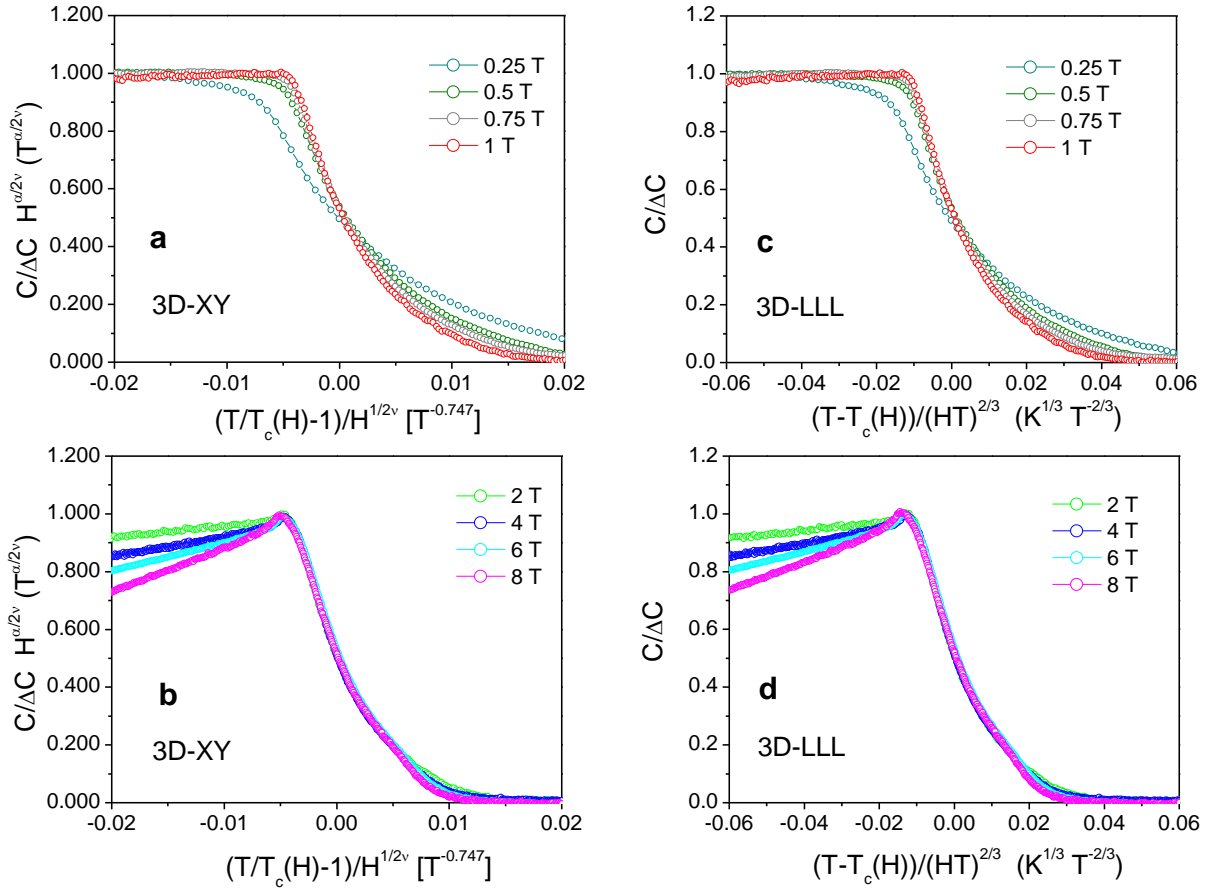

**Figure 2. a, b** 3d-XY and **c, d** 3d-LLL scaling of the superconducting specific heat contribution $C/T-\gamma$ of $V_3Si$ in various applied magnetic fields after separation of the normal-state specific heat. **a** and **c** show the lower field range up to 1 T, demonstrating that the scaling clearly fails for fields below 2 T, while **b** and **d** show the high field range in which good scaling is observed for both models.

## 4. Concluding remarks

With our high-resolution specific heat data we could demonstrate that the observation of fluctuation effects is not limited to HTSCs and $Nb_3Sn$, but are a rather universal feature of classical high-$\kappa$ superconductors with elevated critical temperatures. This information may be essential for a detailed understanding of universal features observed in various type-II superconductors, especially the increasing width of the upper critical field line in magnetic fields [10,12,18,19] and eventually the peak effect, which has been reported to occur in direct vicinity to the regime where we observe strong fluctuations [20,21-24]. Upon comparison of $V_3Si$ with the cuprate HTSC $YBa_2Cu_3O_{7-\delta}$ and the classical superconductor $Nb_3Sn$, it becomes obvious that the fluctuations effects in $V_3Si$ are quite weak. A rough measure of the strength of fluctuations can be taken as the upper boundary of the $x$-axis scaling variable $[T/T_c(H)-1]H^{1/2\nu}$ for which scaling holds. Strong fluctuations appear in $YBa_2Cu_3O_{7-\delta}$ up to $[T/T_c(H)-1]H^{1/2\nu} = 0.1$, for $Nb_3Sn$ up to $[T/T_c(H)-1]H^{1/2\nu} = 0.01$, while for $V_3Si$ a close look on Figure 3c reveals that the scaling starts to fail above $[T/T_c(H)-1]H^{1/2\nu} = 0.006$. This explains that a field of 2 T is needed to widen the critical temperature range around $T_c$ in $V_3Si$ so much that the fluctuations become experimentally observable, while in $Nb_3Sn$ they become already visible in 0.1 T and in $YBa_2Cu_3O_{7-\delta}$ they have been observed even in zero field, up to several tens of Kelvins above $T_c$ [4].

**Acknowledgments**

This research was supported by various grants from the Research Grants Council of the Hong Kong Special Administrative Region, China (603010, SEG_HKUST03 and SRFI11SC02).

**References**

[1] Schneider T, Singer J M 2000 *Phase Transition Approach to High Temperature Superconductivity*, Imperial College Press and references therein.
[2] Roulin M, Junod A, Walker E 1996 *Physica* C **260**, 257-272 (1996).
[3] Roulin M, Junod A, Muller J 1995 *Phys. Rev. Lett.* **75**, 1869.
[4] Meingast C, Pasler V, Nagel P, Rykov A, Tajima S, Olsson P 2001 *Phys. Rev. Lett.* **86**, 1606.
[5] Lortz R, Meingast C, Rykov A I, Tajima S 2003 *Phys. Rev. Lett.* **91**, 207001.
[6] Overend N, Howson M A, Lawrie I D 1994 *Phys. Rev. Lett.* **72**, 3238.
[7] Pierson S W, Katona, T M, Tešanović Z, Valls O T *Phys. Rev.* B **53**, 8638 (1996).
[8] Tešanovic Z 1999 *Phys. Rev.* B **59**, 6449.
[9] Buckingham M J, Fairbank W M 1961 "Chapter III The Nature of the λ-Transition in Liquid Helium". The nature of the λ-transition in liquid helium. Progress in Low Temperature Physics 3. p. 80. doi:10.1016/S0079-6417(08)60134-1.
[10] Farrant S P, Gough C E, 1975 *Phys. Rev. Lett.* **34**, 943.
[11] Thouless D J 1975 *Phys. Rev. Lett.* **34**, 946.
[12] Lortz R, Lin F, Musolino N, Wang Y, Junod A, Rosenstein B, Toyota N 2006 *Phys. Rev.* B **74**, 104502.
[13] Toyota N, Kobayashi T, Kataoka M, Watanabe H F J, Fukase T, Muto Y, Takei F 1988 *J. Phys. Soc. Jpn.* **57**, 3089.
[14] Maki K, 1964 *Physics* **1**, 21.
[15] Zehetmayer M, Hecher J 2014 Supercond. Sci. Technol. **27**, 044006.
[16] Lee P A, Shenoy S R 1972 *Phys. Rev. Lett.* **28**, 1025.
[17] Haussmann R 1999 *Phys. Rev. B* **60**, 12373.
[18] Xiao Z L, Dogru O, Andrei E Y, Shuk P, Greenblatt M 2004 *Phys. Rev. Lett.* **92**, 227004.
[19] Thakur A D, Banerjee S S, Higgins M J, Ramakrishnan S, Grover A K 2005 *Phys. Rev. B* **72,** 134524.
[20] Isino M, Kobayashi T, Toyota N, Fukase T, Muto Y 1988 *Phys. Rev. B* **38**, 4457.

[21] de Sorbo W 1964 *Rev. Mod. Phys.* **36**, 90.
[22] Huxley A D, Paulson C, Laborde O, Tholence J L, Sanchez D, Junod A, Calemczuk R 1993 *J. Phys.: Condens. Matter* **5**, 7709.
[23] Roy S B, Chaddah P 1997 *J. Phys.: Condens. Matter* **9**, L625, Roy S B *et al.* 2000 *Phys. Rev. B* **62**, 9191.
[24] Banerjee S S, Patil N G, Saha S, Ramakrishnan S, Grover A K, Bhattacharya S, Ravikumar G, Mishra P K, Chandrasekhar Rao T V, Sahni V C, Higgins M J, Yamamoto E, Haga Y, Hedo M, Inada Y, Onuki Y 1998 *Phys. Rev. B* **58**, 995.